\documentclass{elsarticle}
\journal{arXiv}

\usepackage{amsmath}
\usepackage{amsfonts}
\usepackage{amssymb}
\usepackage{booktabs}
\usepackage{caption}
\usepackage{graphicx}
\usepackage{float}
\usepackage{fancyref}

\usepackage{graphicx,amssymb,lineno}
\begin{document}
\begin{frontmatter}

\title{How to improve accuracy for DFA technique}

\author{Alessandro Stringhi$^\dagger$}
\author{Silvia Figini$^{\natural}$}

 \address{$^{\dagger}$Department of Physics, University of Pavia, Italy}
 \ead{alessandro.stringhi01@ateneopv.it}
\address{$^{\natural}$Department of Statistics and Applied Economics, University of Pavia, Italy}
                        \ead{silvia.figini@unipv.it}

\begin{abstract}
This paper extends the existing literature on empirical estimation of the confidence intervals associated to the Detrended Fluctuation Analysis (DFA). We used Montecarlo simulation to evaluate the confidence intervals. Varying the parameters in DFA technique, we point out the relationship between those and the standard deviation of $H$. The parameters considered are the finite time length $L$, the number of divisors $d$ used and the values of those. We found that all these parameters play a crucial role, determining the accuracy of the estimation of $H$.

\end{abstract}

\end{frontmatter}

\section{Introduction}
The Hurst exponent $H$ \cite{Hurst} has been applied in several fields and its value is related to specific characteristic of an independent stochastic process. The $H$ value is bounded between $(0,1)$. If $H$ is equal to 0.5, the independent stochastic process doesn't show a long term memory; if $H>0.5$ the series is \emph{persistent} and the process is characterised by a trend reinforcing memory. On the other hand if $H<0.5$ the series is \emph{anti-persistent}.

 According to the Black\&Scholes model and the EMH (Efficient Market Hypothesis), the financial series, such as stock prices and indexes, should display a Hurst exponent equal to $0.5$. This feature has been deeply studied and it is a strong belief that most developed markets show no long-range memory \cite{di_matteo}. The literature underlines that the techniques used to estimate the value of $H$ are sometimes misleading, especially when looking at long time memory with stock market data, the $H$ index is larger than 0.5 \cite{Couillard}.
 This is due to the fact that the $R/S$ analysis estimates $H$ correctly only with an infinite time series \cite{Couillard}. Assuming that long financial time series (5-10 years) display $H=0.5$, hence no long-term memory, it is still possible to study if these series have a short-term memory. For such studies it is necessary to consider a small time window (from some months up to few years) to evaluate the so-called \emph{local} (or \emph{time varying}) Hurst exponent. In these considered periods $H$ can differ significantly from the theoretical value. This leads to a local invalidation of the hyphoteses under the EMH and it allows to use trading techniques to achieve better earnings due to arbitrage oppurtunity \cite{mitra}. 
The analyses that estimate $H$, are associated to an error. Several techniques are proposed in literature to estimate $H$. The $R/S$ and more recently the  \emph{Detrended Fluctuation Analysis} (DFA) improve the efficiency in the $H$ estimation without overestimate $H$ in a finite time length \cite{weron}. We point out that the distribution of the DFA is not known and the confidence intervals of the measure have to be calculated using a Montecarlo simulation \cite{weron} \cite{ladislav}.

The main aim of this paper is to explore the behaviour of the $H$ index estimation using the DFA technique with a special focus of the evaluation of confidence intervals which are estimated with a better precision using 40'000 points and considering very small time lengths (starting from 60 days), couple with a sensitivity analysis to understand how changement in the parameters affects the precision of the results varying the time length ($L$), the number of divisors of $L$ ($d$) and the best selection of those.

The paper is structured as follow: Section 2 introduces the \emph{Detrended Fluctuation Analysis}; Section 3 shows an empirical data analysis based on simulated data underlying the dependence on the $H$ estimates with respect to the parameters involved in the analysis and section 4 describes the conclusions and propose further ideas of research. 


\section{Detrended Fluctuation Analysis (DFA)}

One of hte method used to measure the long-range dependence in data series is the \emph{Detrended Fluctuation Analysis (DFA)} proposed by Peng et al. \cite{PhysRevE.49.1685}. Let $S_t$, $t=1,\dots, T$ be a financial time series; First we consider the log-returns, $r_t=\ln S_t - \ln S_{t-1}$. After dividing the time series into a subset of length $L<T$, we construct the cumulative time series
\begin{equation}
X(t)=\sum_{t=1}^L (r_t-\bar{r}),
\end{equation}
where $\bar{r}$ is the mean value of the data (or log-returns) $r_t$. Second, we divide the cumulative series into $d$ disjoint subseries of length $N_i,\  i=1,\dots,d$. Each $N_i$ has to be a divisor of $L$. For each subseries it is necessary to compute the linear trend function $Y_i(t)$ which fits the cumulative data using the least square estimation. In order to estimate the Hurst exponent we have to introduce the fluctuation function, defined as the standard deviation of the detrended signal:
\begin{equation}
F(N_i)=\sqrt{\frac{1}{L} \sum_{t=1}^L [X(t)-Y_{i}(t)]^2} \quad i=1,\dots,d.
\end{equation}
The fluctuation function is related to $H$ according to the law $F(N)\propto N^H$, thus plotting in a log-log scale $F(N)$ with respect to $N$ we estimate $H$ as a angular coefficient of the linear trend function.
If the considered time series does not display a long-range memory, the estimated value of $H$ has to be equal to $0.5$. If the value found is $H>0.5$, it means that the series is persistent, otherwise if $H<0.5$, the series is anti-persistent.  

To our knowledge no asymptotic distribution for the \emph{DFA} is known. This lead us to investigate how to find confidence intervals associated to the DFA technique for $H$ estimation using Montecarlo simulation.

\section{Empirical Analysis}



The estimation of the Hurst exponent is made using simulated data drawn from a standardized normal distribution; the sample size is equal at 210000 observations. $H$ is estimated using two types of length:
\begin{itemize}
\item using the power of 2 (case A); 
\item using the multiples of 60 (case B). 
\end{itemize}
The case B is appealing because presents the greatest number of divisors among the integer numbers close to $2^n$. In our analysis concerning the case B, not all the possible divisors has been used; we limit our analysis using a number of divisors double with respect to case A. In both cases we used divisors $\geq 8$. Table 1 reports the parameters setting for the two analysis. In table 1 $L$ is the time length considered, $d$ is the number of divisor used and $N_{min}$ is the smallest divisor among the $d$ divisors considered.
\\

\begin{table}[ht]
\caption{Parameter setting}
\centering
\begin{tabular}{cccc}
\toprule
\multicolumn{2}{c}{case A} & \multicolumn{2}{c}{case B}\\
\midrule
$L(d)$ & $N_{min}$ & $L(d)$ & $N_{min}$\\
\midrule
64 (3) & 8 & 60 (5) & 10\\
128 (4) & 8 & 120 (8) & 10\\
256 (5) & 8 & 240 (10) & 12\\
512 (6) & 8 & 480 (12) & 15\\
1024 (7) & 8 & 960 (14) & 20\\
2048 (8) & 8 & 1920 (16) & 30\\
4096 (9) & 8 & 3840 (18) & 40\\
8192 (10) & 8 & 7680 (20) & 48\\
\bottomrule
\end{tabular}
\end{table}


In table 2 and table 3 we report the mean and the standard deviation of the 40000 values of $H$.
\begin{table}[ht]
\caption{Case A}
\centering
\begin{tabular}{ccccccccc}
\toprule
L	& 64 & 128 & 256 & 512 & 1024 & 2048 & 4096 & 8196\\
\midrule
mean & 0.4991 & 0.4961 & 0.4954 & 0.4974 & 0.4997 & 0.5020 & 0.5010 & 0.4977\\
SD 	& 0.1549 & 0.1033 & 0.0752 & 0.0587 & 0.0467 & 0.0377 & 0.0290 & 0.0237\\
\bottomrule
\end{tabular}

\end{table}

\begin{table}[ht]
\caption{Case B}
\centering
\begin{tabular}{ccccccccc}
\toprule
L	& 60 & 120 & 240 & 480 & 960 & 1920 & 3840 & 7680\\
\midrule
mean & 0.4903 & 0.4902 & 0.4884 & 0.4916 & 0.4954 & 0.4994 & 0.5028 & 0.4994\\
SD 	& 0.1842 & 0.1161 & 0.0894 & 0.0753 & 0.0627 & 0.0543 & 0.0437 & 0.0341\\
\bottomrule
\end{tabular}
\end{table}
The empirical evidence shows that in both cases $H$ is not overestimated for each $L$ ( as shown in \cite{weron} \cite{ladislav} using the $R/S$ analysis ). Table 2 and 3 for each $L$ considered in case A and B depict a mean constant around the asymptotic value of $0.5$ and a standard deviation which decrease when $L$ increase. Comparing the two cases we note that case A has a lower standard deviation associated to the measure with respect to the case B.

\begin{figure}[H]
\centering
\includegraphics[scale=0.5]{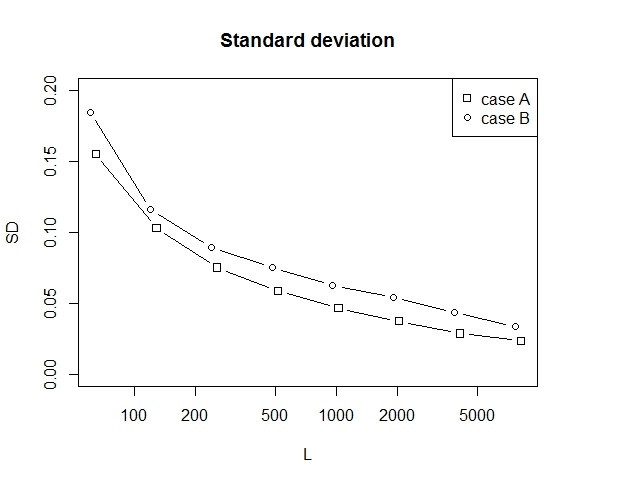}
\caption{Standard deviation for case A and case B}
\label{fig:stdev}
\end{figure}

Figure \ref{fig:stdev} reports the comparison between the standard deviation obtained in case A and case B. Notice that in figure \ref{fig:stdev} the behaviour of the two curves appears not so intuitive because we expect that curve A should be upper the curve linked to case B due to the less number of divisors.

One the basis of the data used described before, confidence intervals are built using Montecarlo simulation. The confidence levels are derived using the 3$\sigma$-law, without resorting to the Gaussian assumption. The confidence levels are 68.3\%, 95.5\% and 99.7\% respectively.

\begin{figure}[H]
\centering
\includegraphics[scale=0.45]{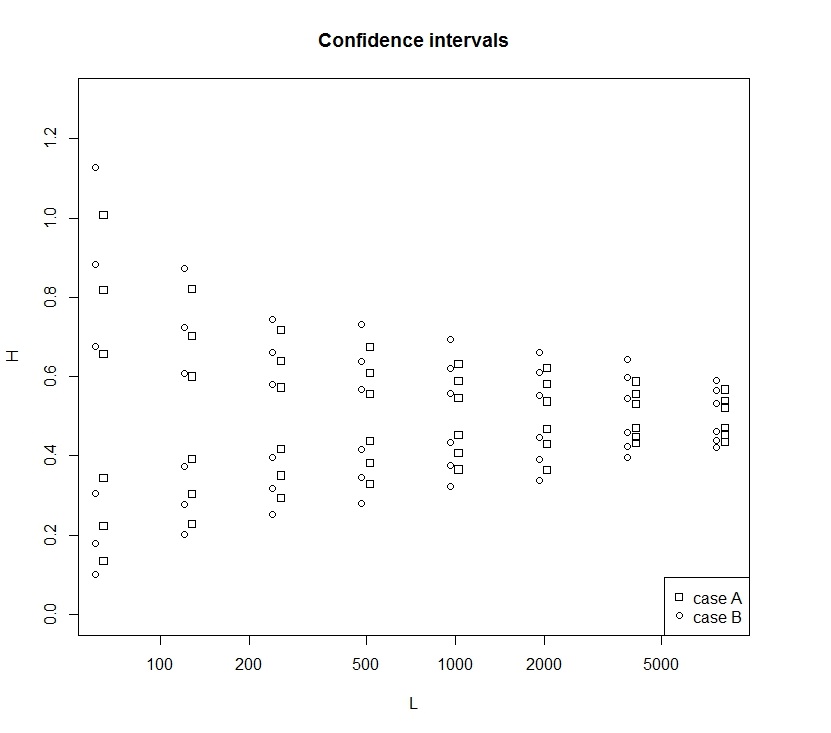}
\caption{Confidence intervals for case A and case B}
\label{fig:cl}
\end{figure}

Figure \ref{fig:cl} plots the confidence intervals obtained for case A and case B. As we can observe from figure \ref{fig:cl} if $L$ increase, the confidence bounds are close. On the other hand, low level of $L$ exhibit large intervals. From figure \ref{fig:cl} we remark that the confidence intervals associated to case B are wider with respect to the confidence intervals obtained in case A.


Is well known that DFA is more accurate when a big number of divisors are available. This is motivated to the fact that $H$ is estimated with a linear fit between $F(N)$ and $N$, in a log-log scale plot. Thus a linear fit is more accurate with a greater number of points. As previously shown in figure \ref{fig:stdev} and figure \ref{fig:cl}, case A is more accurate despite the lower number of divisor, the half respect case B. For verifying the aforementioned hypothesis we estimated 10000 values of $H$ with a Montecarlo simulation on normal distributed data. We used $L=3840$ and a number of divisor $d$ from 6 to 18.

\begin{figure}[H]
\centering
\includegraphics[scale=0.4]{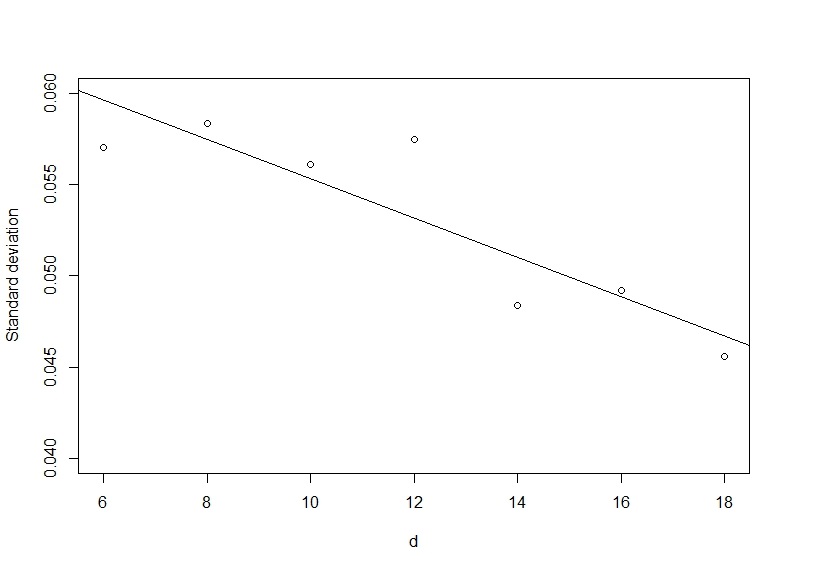}
\caption{Value of standard deviation respect the number of divisor}
\label{fig:div}
\end{figure}

The analysis confirms the hypothesis that the DFA accuracy depends on the number of divisor $d$ used in the analysis. The equation of the fit line is $S=-0.001d+0.066$ with $R^2=0.787$. According to the formula the standard deviation associated to $L=2^{12}=4096$ should be approximately 0.6, but the experimental value found is 0.0290. The reason of the more accuracy related to the case A should be investigated elsewhere.


On the basis of the results at hand we find that the number of divisor is an important parameter to achieve a better accuracy, but despite that case A is still more accurate than case B. But Returning back to the first analysis we can notice that in case A we used all divisors available, from 8 to $2^{d-1}$, while in case B we choose to discharge the lowest divisors. Divisor selection is the real problem to be solved. In this last analysis we simulated again 40'000 value of $H$, using the same data used in the first analysis. We also set $L=1920$ and $d=8$, but we choose 5 different set of consecutive divisors, from the lowest to the highest.

 \begin{figure}[H]
\centering
\includegraphics[scale=0.4]{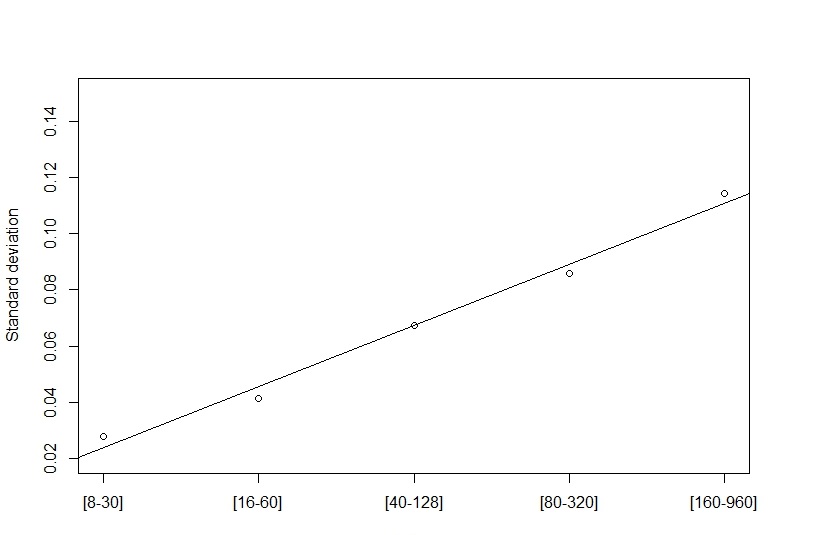}
\caption{Value of standard deviation respect the choice of the divisor}
\label{choice}
\end{figure}

The graph in figure \ref{choice} shows how, keeping $L$ and $d$ constant, the right choice of the divisor used in the DFA influence the standard deviation of the measure. Using the lowest divisor available allow to reduce the standard deviation in a significant way. Considering the standard deviation associated to the lowest divisor we find that, for case B ($L=1920$) $S=0.0279$ and for case A ($L=2048$) $S=0.0377$, in both cases $d=8$. On the basis of the empirical evidence at hand the choice among all divisors available is crucial for DFA.

\section{Conclusions} 
 
The multi-fractional Brownian motion is a stochastic process with wide and important applications. A proper knowledge about the techniques used to estimate the Hurst exponent is necessary to avoid misleading results. In this paper we have compared two different set of time length, based on the power of 2 and the multiples of 60. More precisely the final aim is to derive the choice of $L$ leading to and estimate of $H$ more accurate. Case A shows a lower standard deviation and more narrow confidence intervals with respect to case B, despite case B uses a double number of divisor $d$. We also showed the dependence between the standard deviation and some important parameter of the DFA technique. The number and the right choice of the divisor influence deeply the accuracy of the measure. We pointed out how increasing the number of divisor $d$ the standard deviation decrease, this result was expected because in DFA the Hurst exponent is estimated using a linear fit between the fluctuation function $F(N)$ and $N$, so using more point,  conversely more divisor, the fit line is more accurate and respectively the estimation of $H$. Less intuitive is the dependence of the standard deviation respect to the choice of the divisors. Using the lowest available divisor we found that the standard deviation can be reduced even by a factor 4. Despite this evidence we still feel to reccomend to avoid to use divisor lower than 8. These parameter also influence the computational time of the DFA. Sometimes the computational time can be a crucial factor, but it has not been deeply investigated in this paper. Nevertheless looking at the code used for the DFA is possible to understand how the parameters influence the computational time. It result longer when we perform the analysis with an increasing $L$ or $d$. Instead the computational time results longer when low divisors are used. We think that this approach may be interesting for practitioners especially in financial time series analysis.

\bibliographystyle{plain}
\bibliography{bibliografia}

\end{document}